\documentclass[english,aps,reprint]{revtex4}
\usepackage[T1]{fontenc}
\usepackage[latin9]{inputenc}
\usepackage{babel}
\usepackage{float}
\usepackage{textcomp}
\usepackage{amstext}
\usepackage{graphicx}
\usepackage{subscript}
\usepackage[unicode=true]
 {hyperref}

\makeatletter

\providecommand{\tabularnewline}{\\}

\@ifundefined{textcolor}{}
{%
 \definecolor{BLACK}{gray}{0}
 \definecolor{WHITE}{gray}{1}
 \definecolor{RED}{rgb}{1,0,0}
 \definecolor{GREEN}{rgb}{0,1,0}
 \definecolor{BLUE}{rgb}{0,0,1}
 \definecolor{CYAN}{cmyk}{1,0,0,0}
 \definecolor{MAGENTA}{cmyk}{0,1,0,0}
 \definecolor{YELLOW}{cmyk}{0,0,1,0}
 }

\makeatother

\begin{document}

\title{Observation of a rotational transition in trapped and sympathetically
cooled molecular ions}

\author{J. Shen, A. Borodin, M. Hansen, S. Schiller}

\address{Institut für Experimentalphysik, Heinrich-Heine-Universität Düsseldorf,
40225 Düsseldorf, Germany}
\begin{abstract}
We demonstrate rotational excitation of molecular ions that are sympathetically
cooled by laser-cooled atomic ions to a temperature as low as ca.~10~mK.
The molecular hydrogen ions HD\textsuperscript{+} and the fundamental
rotational transition $(v=0,\, N=0)\rightarrow(v'=0,\, N'=1)$ at
1.3 THz, the most fundamental dipole-allowed rotational transition
of any molecule, are used as a test case. This transition is here
observed for the first time directly. Rotational laser cooling was
employed in order to increase the signal, and resonance-enhanced multiphoton
dissociation was used as detection method. The black-body-radiation-induced
rotational excitation is also observed. The extension of the method
to other molecular species is briefly discussed.
\end{abstract}
\maketitle

\section{Introduction}

High-resolution laboratory rotational spectroscopy of molecules is
an important and very well developed technique in molecular physics.
It has provided extensive data on and insight into the structure and
dynamics of molecules and has found, among other, application for
the identification of molecular species in interstellar clouds. In
the recent past, the accessible spectral region has been extended
from the microwave region to the terahertz (THz, sub-millimeter wavelength)
region, thanks to the development of appropriate THz radiation sources
attaining useful power levels. Continuous-wave, narrow-linewidth THz
radiation, suitable for high-resolution molecular spectroscopy, is
available from backward-wave oscillators or via frequency upconversion
using Schottky diodes \cite{Lewen,Maestrini} or semiconductor superlattices
\cite{Klappenberger,Endres}.

The resolution of rotational spectroscopy has been increased beyond
the Doppler limit by several techniques, such as molecular beams \cite{Kroto},
velocity-class selection \cite{Carocci}, Lamb-dip spectroscopy, and
two-photon spectroscopy \cite{Surin}. For example, sub-Doppler spectral
lines with widths of ca. 15 kHz around 100 GHz \cite{Cazzoli} and
several 10~kHz at 0.7 - 1 THz \cite{Winnewisser 1997,Ahrens} have
been reported. The resolution of such methods is, however, limited
by transition time broadening and/or pressure broadening. In order
to eventually overcome these limitations, it is interesting to explore
a fundamentally different regime: trapping cold (< 1 K) molecules
in an interaction-free (ultra-high-vacuum) environment and localization
to sub-mm extension. Then, transition time broadening and pressure
broadening, as well as Doppler broadening may be strongly reduced
or even eliminated altogether. This regime may be reached with cold
neutral molecules stored in optical traps or with cold molecular ions
in radio-frequency traps. The production methods of cold molecules
have been reviewed elsewhere, see e.g. the references \cite{Cold Molecules book,Cold Molecules book2}.

In this work, we take a first step towards applying high-resolution
rotational spectroscopy on cold molecules: we demonstrate, for the
first time to our knowledge, rotational excitation on sympathetically
cooled and strongly confined molecular ions, using here temperatures
as low as 10-15~mK. Our spectroscopic technique is a destructive
one, resonance-enhanced multiphoton dissociation (REMPD). The particular
implementation used here is 1+1'+1'' REMPD, where the molecule absorbs
sequentially three photons of different energy. In our work, the transitions
induced by each photon corresponds to the three main energy scales
of a molecule: rotational (here ca. 1 THz), vibrational (here: overtone,
ca. 200 THz), and electronic (ca. 1100 THz). 

With respect to the temperature of the ions and the detection method,
our work is complementary of the recent demonstration of THz rotational
spectroscopy of helium buffer gas cooled H\textsubscript{2}D\textsuperscript{+}and
D\textsubscript{2}H\textsuperscript{+} ions at kinetic ion temperatures
of ca. 24~K, where Doppler broadening is limiting the linewidth (to
ca. 1~MHz) \cite{Asvany}.

\section{The HD\textsuperscript{+}molecule and the spectroscopic technique}

The molecule we consider here, HD$^{+}$, is the most fundamental
molecule with electric dipole-allowed rotational transitions \cite{Wing}.
Its potential as a test system for molecular quantum mechanics and
for novel fundamental physics studies has been described previously
\cite{Wing,Carrington 1989,Schiller and Korobov 2005,Korobov 2006,Korobov 2007,Koelemeij 2007}.
Vibrational spectroscopy of sympathetically cooled HD$^{+}$ with
the highest resolution of any molecular ion to date has recently been
reported by us \cite{Bressel 2011}. Pure rotational transitions have
so far been observed only for the last and penultimate vibrational
levels, $v=21,\,22$, close to the dissociation limit \cite{Carrington - pure rotational v 21,Carrington - pure rotational v 22},
where their $N=0\rightarrow N'=1$ transition frequencies lie in the
microwave range (ca.~50~GHz and 9.4~GHz, respectively). Here we
report on the rotational transition in the $v=0$ ground vibrational
level, at much higher frequency. The ro-vibrational transition frequencies
have been calculated \textit{ab-initio} with high precision \cite{Korobov 2006,Korobov 2007},
greatly facilitating the experimental search. The $(v=0,\, N=0)\rightarrow(v'=0,\, N'=1)$
fundamental rotational transition studied here occurs at $f_{0,th}=$
1~314~925.752~MHz ({}``spinless'' value, i.e. excluding hyperfine
energy contributions), with an estimated theoretical error of ca.
2 kHz \cite{Korobovprivcomm}. The $f_{0,th}$ value includes a contribution
of ca. 48.8 MHz from relativistic effects (order $\alpha^{2}$) and
ca. -9.4 MHz from QED effects of order $\alpha^{3}$.

\begin{figure}[H]
\begin{centering}
\includegraphics[width=11cm]{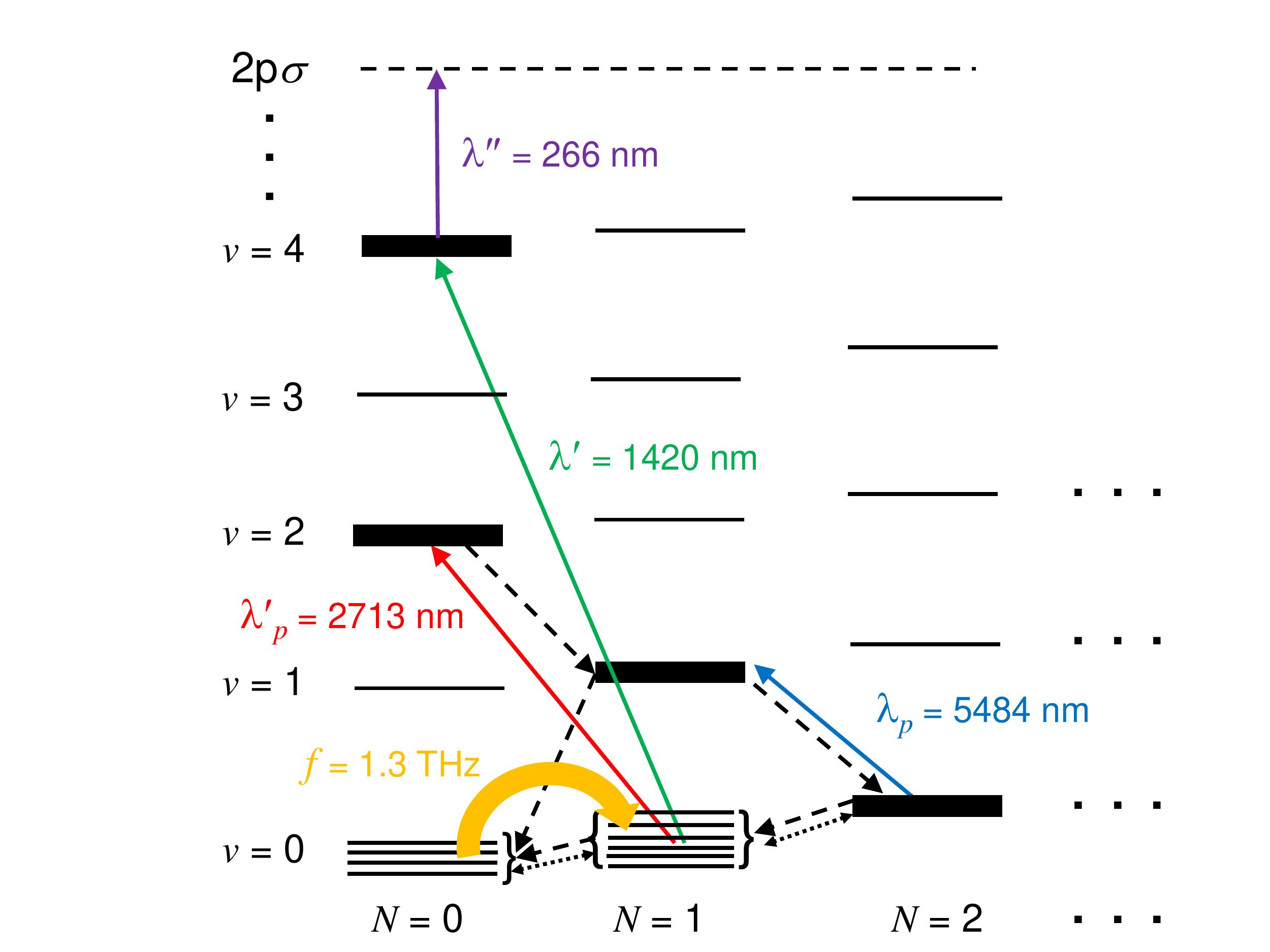}
\par\end{centering}

\caption{\label{fig:Schematic-energy-level} \textbf{Simplified energy level
scheme of HD$^{+}$ with transitions} \textbf{relevant to this work}.
Full thin arrows: laser-induced transitions, dashed arrows: some relevant
spontaneous emission transitions; dotted double arrows: some relevant
black-body-radiation-induced transitions. The THz wave (thick arrow)
is tuned so that the four hyperfine states in $(v=0,\, N=0)$ are
excited to corresponding hyperfine states in $(v'=0,\, N'=1).$ Resonant
laser radiation at $\lambda'$ (1420 nm) and non-resonant radiation
at $\lambda''$ (266 nm) transfer the rotationally excited molecules
to a vibrationally excited level $(v''=4,\, N''=0)$ and then further
to electronically excited molecular states (predominantly 2p$\sigma$),
from which they dissociate. Rotational cooling is performed by radiation
at $\lambda_{p}$ (5.5 \textmu{}m) and $\lambda_{p}'$ (2.7 \textmu{}m).
The level energy differences are not to scale. Hyperfine structure
is indicated very schematically for the levels $(v=0,\, N=0,\,1)$
and as thick lines for some other participating levels. The waves
at $\lambda',\,\lambda'',\,\lambda_{p},\,\lambda_{p}'$ have relatively
large spectral linewidths.}
\end{figure}

An important aspect of this work is the use of laser rotational cooling
\cite{Schneider 2009}, see Fig.~\ref{fig:Schematic-energy-level}.
It is used to transfer most molecular population, initially distributed
among several rotational levels in $v=0$, into the ro-vibrational
ground state $(v=0,\, N=0)$ and it also empties the spectroscopy
target state $(v'=0,\, N'=1).$ It modifies the difference in fractional
population of the lower and upper spectroscopy levels, from ca. -0.15
in thermal equilibrium to ca. 0.7, and thus significantly increases
the detectable signal.

\begin{figure}
\centering{}\includegraphics[scale=0.5]{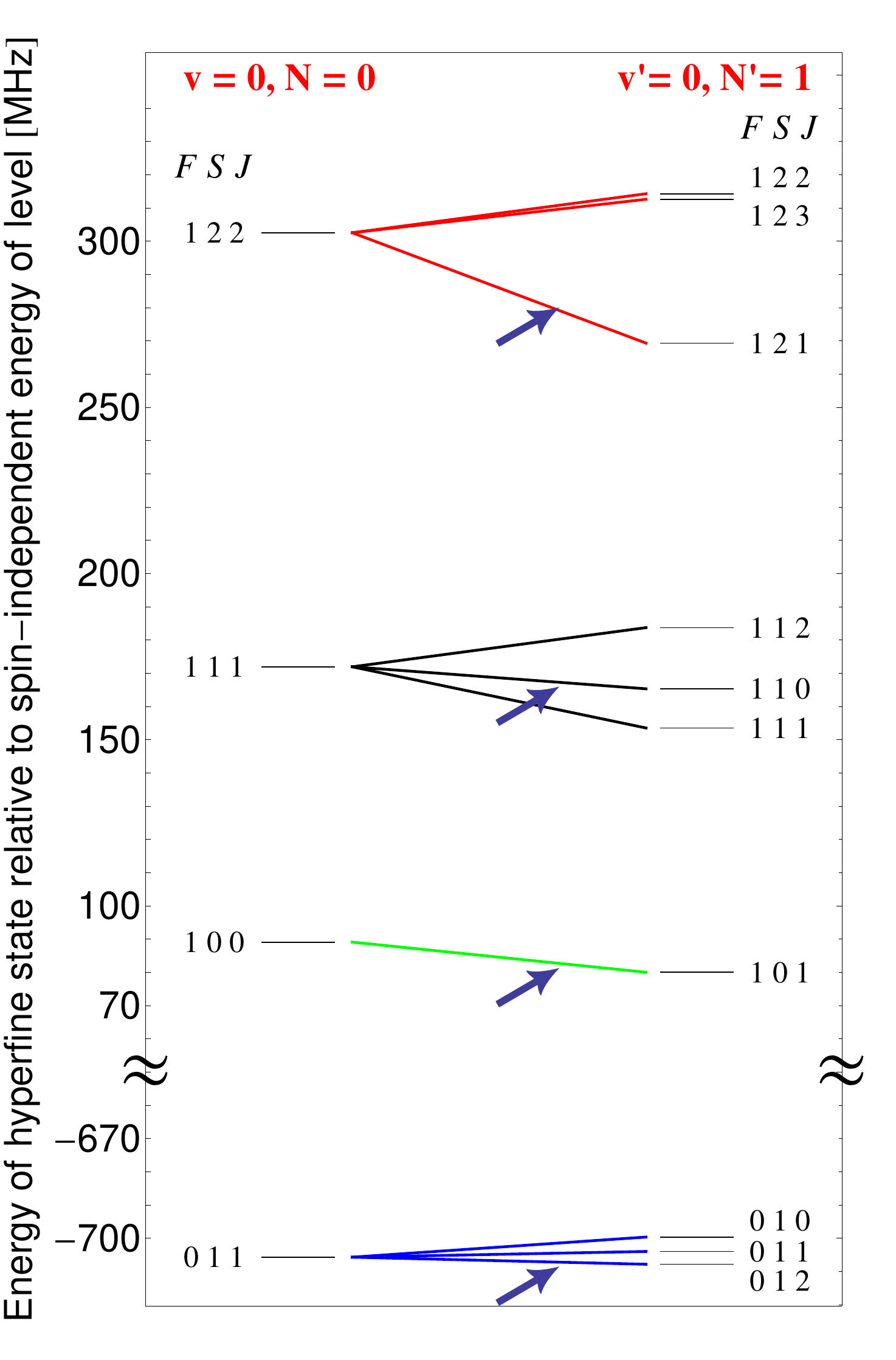}\caption{\textbf{\label{fig:Energy-diagram-of}Energy diagram of the hyperfine
states and main electric-dipole transitions in zero magnetic field.}
Left side: ro-vibrational ground level $(v=0,\, N=0)$, right side:
rotationally excited level $(v'=0,\, N'=1)$. The hyperfine states
are labeled by the (in part approximate) quantum numbers $(F,\, S,\, J)$.
The degeneracy factor of each hyperfine state is $(2J+1)$. Transitions
that do not change the quantum numbers $\ensuremath{F,\, S}$ are
relatively strong and are indicated by red, black, green, and blue
lines. The four transitions addressed simultaneously in this work
are indicated by arrows.}
\end{figure}
\begin{figure}
\centering{}\includegraphics[scale=0.55]{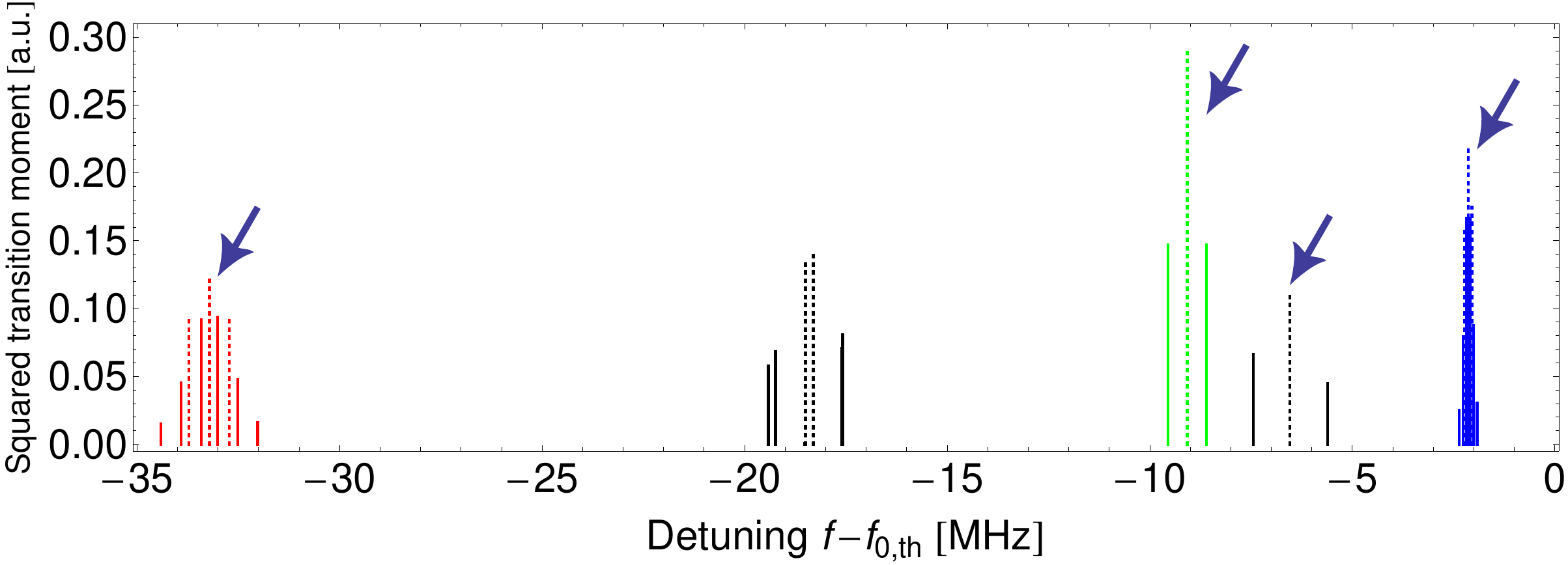}\caption{\label{fig: Stick spectrum} \textbf{Section of the theoretical stick
spectrum of the rotational transition in the frequency range relevant
to this work, showing the Zeeman splittings and shifts in a 1 G magnetic
field.} Dashed lines are $\pi$ transitions, full lines are $\sigma$
transitions. $f_{0,th}$ is the theoretical {}``spinless'' transition
frequency. The four THz frequencies of list A (see Table \ref{tab: Table})
are indicated by the arrows. The colors used correspond to those used
in Fig.~\ref{fig:Energy-diagram-of}.}
\end{figure}

The hyperfine structure and the Zeeman effect of the lower and upper
ro-vibrational levels are important aspects in the rotational spectroscopy
\cite{Bakalov 2006,Bakalov 2010b}. As Fig.~\ref{fig:Energy-diagram-of}
shows, the ground state possesses four hyperfine states and altogether
12 magnetic substates (with magnetic quantum number $J_{z}$). In
the ions' region the magnetic field is nonzero, lifting the magnetic
degeneracy, The spectrum contains a large number of transitions with
relatively large transition dipole moments, see Fig.~\ref{fig: Stick spectrum}.
Most transitions, even the $\Delta J_{z}=0$ ones, shift by on the
order of 100 kHz or more in a field of strength 1 G. Exceptions include
five (strong) $J_{z}=0\rightarrow J_{z}'=0$ transitions, whose quadratic
Zeeman shifts in 1 G are at most 6.2 kHz in absolute value \cite{Bakalov 2010b}.
From each lower hyperfine state there is at least one such transition;
the state $(F=0,\, S=1,\, J=1,\, J_{z}=0)$ has two. Three of them
are indicated by the first, third, and fourth arrow (from the top)
in Fig.~\ref{fig:Energy-diagram-of}. A fourth is the hyperfine transition
$(F=1,\, S=1,\, J=1,\, J_{z}=0)$~$\rightarrow(F'=1,\, S'=1,\, J'=2,\, J_{z}'=0)$.
Its transition frequency $f=f_{0,th}+11.78$~MHz is close to other
transition frequencies and is therefore not considered suitable for
the present work. Instead, we use the $(F=1,\, S=1,\, J=1,\, J_{z}=0)$~$\rightarrow(F'=1,\, S'=1,\, J'=0,\, J_{z}'=0)$.
transition (second arrow in Fig.~\ref{fig:Energy-diagram-of}), which,
however, exhibits a much larger quadratic Zeeman effect, e.g. 78 kHz
at 1 G. The magnetic field in the trap region is not spatially constant
in direction and magnitude, so some line broadening can be expected. 

The theoretical Doppler linewidth is ca. 55 - 70 kHz for the lowest
temperatures used here (10 - 15 mK) and ca. 150 - 200 kHz if the ion
ensemble is in the liquid state (100 - 200 mK). Note that the values
are relatively large due to the low mass of the HD\textsuperscript{+}
ion. At 10 - 15 mK the molecular ions are well confined along the
axis of the trap and their motion in transverse direction is restricted
to a range smaller than the THz wavelength (0.23 mm). In the axial
direction, the confinement is not as strong, since diffusive motion
of the ions is still occurring along the crystal axis, which exceeds
in length (ca. 1.5 mm) the THz wavelength. Thus, Doppler broadening
may still be present in our experiment, even at the lowest temperatures.
The THz source has a linewidth below 100 Hz and sub-Hz absolute frequency
stability, values that are not relevant in comparison to other line
broadening effects. 

The Doppler linewidth is smaller than the typical spacing between
hyperfine transitions originating from different ground hyperfine
states, even in presence of a magnetic field on the order of 1 G.
This is in principle advantageous, since it could permit to resolve
the hyperfine structure. However, detecting individual hyperfine lines
would also lead to a small signal-to-noise ratio: the individual hyperfine
levels each contain only a fraction of the total population. If a
statistical population distribution were produced by action of the
black-body radiation field (BBR) and of the rotational cooling lasers,
each substate would contain 1/12 of the total population, this fraction
being typically 30 molecules. The individual populations are likely
to vary significantly in time, requiring the collection of substantial
data and averaging. We do not attempt to do so here and in order to
obtain a sufficiently strong rotational excitation signal we have
applied the following strategy. 

We irradiate the molecules sequentially on THz frequencies corresponding
to strong hyperfine transitions. Different frequency sets are used,
see Table \ref{tab: Table}. Each frequency from a set is irradiated
for 200 ms, and is meanwhile frequency-modulated by $\pm2\,$kHz at
a 5~Hz rate. The list is repeated several times for a total of 3~s
or more, depending of the excitation approach used.
\begin{table}
\begin{tabular}{|c|c|c|c|c|c|c|c|}
\hline 
~Lower hyperfine level $(F,\, S,\, J)$~  & (1,2,2) & \multicolumn{4}{c|}{(1,1,1)} & (1,0,0) & (0,1,1)\tabularnewline
\hline 
 & \multicolumn{7}{c|}{Frequency $f_{i}-f_{0,th}$ {[}MHz{]} }\tabularnewline
\hline 
List A' & -33.211 & -6.597 & -6.578 & -6.558 & -6.539 & -9.069 & -2.138\tabularnewline
\hline 
List A & -33.211 & \multicolumn{4}{c|}{-6.539} & -9.069 & -2.138\tabularnewline
\hline 
List B & -34.993 & \multicolumn{4}{c|}{-7.850} & -9.773 & -2.465\tabularnewline
\hline 
List C & -31.408 & \multicolumn{4}{c|}{-5.096} & -8.355 & -1.812\tabularnewline
\hline 
List D & -34.102 & \multicolumn{4}{c|}{-7.194} & -9.421 & -2.301\tabularnewline
\hline 
List E & -32.310 & \multicolumn{4}{c|}{-5.817 } & -8.712 & -1.975\tabularnewline
\hline 
{}``500 MHz detuning'' & \multicolumn{7}{c|}{ 500}\tabularnewline
\hline 
\end{tabular}\caption{\label{tab: Table}\textbf{Frequency lists used for excitation of
the rotational transition.} $f_{i}$ is the THz frequency.}
\end{table}

The frequency lists include three of the five above mentioned low-Zeeman-shift
$J_{z}=0\rightarrow J_{z}'=0$ transitions, which originate from three
of the four hyperfine states of $(v=0,\, N=0)$, and the respective
frequency values have been chosen to be those corresponding to an
assumed magnetic field of 1 G. As mentioned above, the chosen transition
starting from the fourth hyperfine state, $(F=1,\, S=1,\, J=1)$ has
a substantial quadratic Zeeman effect. We therefore attempt to compensate
for the lack of precise knowledge of the magnetic field by exciting
at four distinct frequencies, corresponding to the shifts induced
by the magnetic field values (0.25, 0.5, 0.75, 1)~G (the detuning
for 0~G is -6.617~MHz). Altogether, this list of frequencies (denoted
by A', see Table \ref{tab: Table}) should nominally excite four of
the twelve Zeeman substates. However, in the presence of a significant
Doppler broadening, more substates (with larger Zeeman shifts) will
be addressed. Indeed, the frequencies necessary to excite all Zeeman
substates of the ground hyperfine states fall into ranges of ca. $(\pm0.5\,$,
($\pm1\,$), - , $\pm0.22)\,$MHz at 1~G, relative to the frequencies
of the list A'. These spreads have a partial overlap with the Doppler
broadening at the higher molecular temperatures (100 - 200 mK) employed
here.

In order to obtain information about the detuning dependence of the
rotational excitation, we also apply {}``detuned-frequency lists''.
Lists labeled B and C are detuned to smaller and larger frequencies,
respectively, relative to list A' (detunings of list B: $(-1.782,-1.311,-0.704,-0.327)$
MHz, detunings of list C: $(1.803,\,1.443,\,0.714,\,0.326)$ MHz).
The detunings are larger than the shift induced by the Zeeman effect
in a 1~G field. 

Lists D, E have detunings approximately half as large as those of
lists B, C, respectively (list D: $(-0.891,-0.655,-0.352,-0.163)$
MHz and list E: $(0.901,\,0.722,\,0.357,\,0.163)$ MHz). 

Finally, a list A is also used, which is a simplified version of list
A'. 

Frequency modulation is used in all cases.

\section{Experimental apparatus and procedure}

\begin{figure}
\centering{}\includegraphics[bb=0bp 0bp 1134bp 907bp,clip,angle=90,origin=c,scale=0.35]{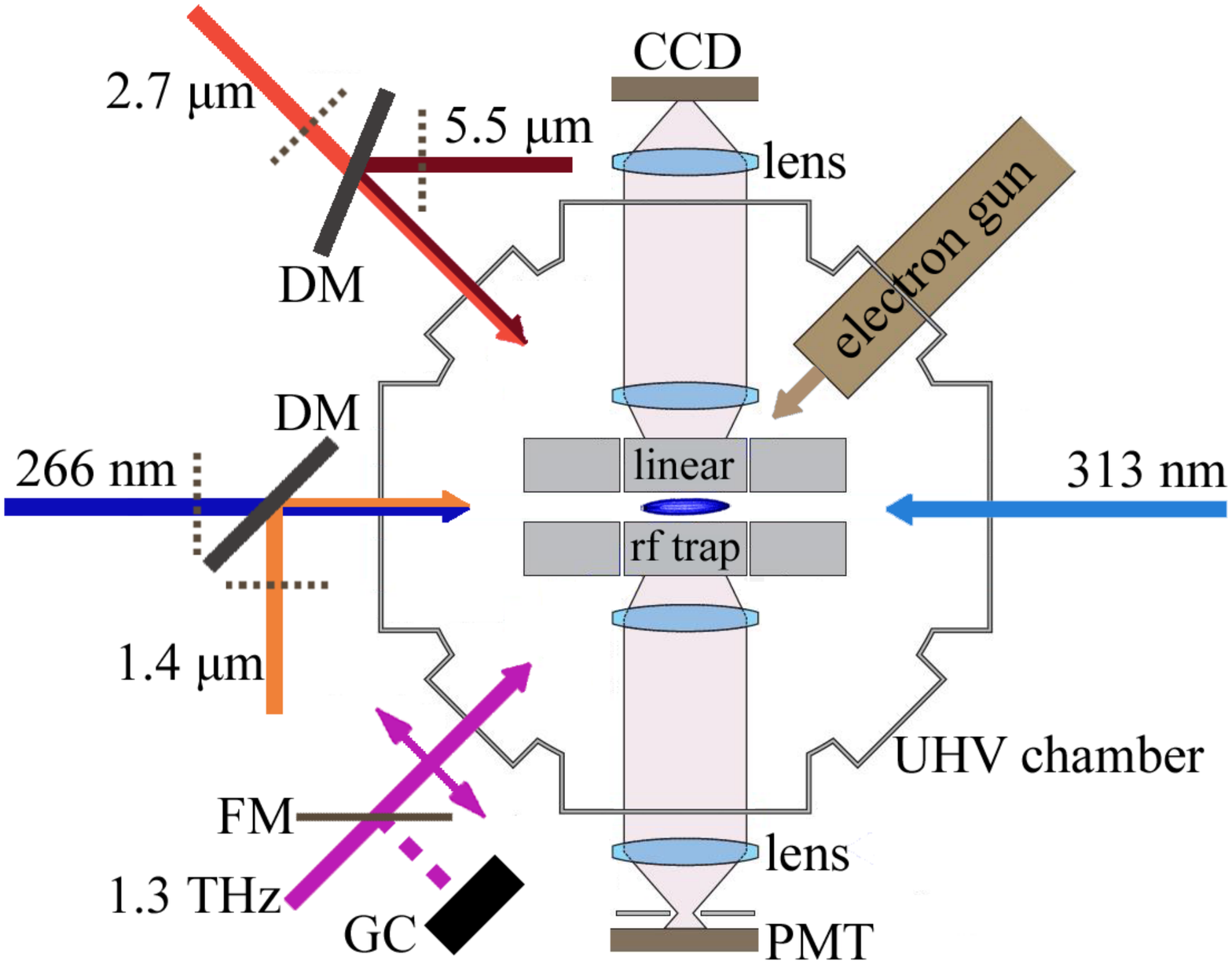}\caption{\label{fig: Layout of apparatus}\textbf{Schematic of the apparatus
and beams. }GC: Golay cell; FM: flip mirror; DM: dichroic mirror;
dotted lines: electrically controlled laser beam shutters. The double
arrow indicates the polarization of the THz wave. Not to scale.}
\end{figure}

A schematic of our apparatus \cite{Blythe 2005} is shown in Fig.~\ref{fig: Layout of apparatus}.
The UHV chamber houses a linear ion trap driven at 14.2 MHz. HD gas
is loaded into the chamber by opening a piezoelectric valve, after
which it is ionized by an electron gun. The laser radiation for REMPD
enters from the left, the 313 nm cooling radiation \cite{Vasilyev 2011}
from the right. The rotational cooling radiation enters diagonally.
The THz source is positioned close to the vacuum chamber and its wave
is focused into the chamber center by a concave paraboloidal mirror
oriented at right angle to the beam. With a manual flip mirror, the
wave can be sent to a Golay cell detector for power monitoring purposes.
The THz source has been described previously \cite{Schiller THz}.
It is driven at the 72$^{nd}$ subharmonic of the desired frequency,
near 18 GHz. This signal is provided by a microwave synthesizer, which
is frequency-locked to a to a GPS-referenced hydrogen maser. The first
REMPD laser is a diode laser emitting at the wavelength $\lambda=1420\,$nm
and exciting the $(v=0,\, N=1)\rightarrow$ $(v'=4,\, N'=0)$ transition.
The second REMPD laser excitation is non-resonant and is provided
by a resonantly frequency-doubled 532~nm laser. Initially, a Be$^{+}$
ion crystal is produced in the trap. Typically, the same Be\textsuperscript{+}
ion crystal is used for several hours of experimentation. 

The molecular sample preparation routine starts by frequency-stabilization
of the cooling laser to a frequency a few tens of MHz to the red of
the frequency for optimal Be$^{+}$ cooling, using a hyperfine transition
of molecular iodine as reference. Then, a small amount of HD gas is
let into the chamber and ionized by the electron gun. Both HD$^{+}$
ions as well as heavy impurity ions are generated, trapped and quickly
sympathetically cooled. In order to remove these impurity ions, the
dc quadrupole potential is briefly increased, reducing the quasi-potential
strength in one transverse direction. The heavy ions then escape from
the trapping region. This ends the preparation procedure; the produced
cold HD$^{+}$ sample contains typically ca. 300 molecules. 

The acquisition of one data point proceeds as follows: (i) the THz
excitation is initially effectively kept off by detuning the frequency
by 500 MHz from the rotational resonance, and the REMPD lasers are
off as well; (ii) the rotational cooling laser beams are unblocked,
and a repeated secular frequency scan (740 - 900 kHz) is activated
and kept on during the remainder of the measurement cycle (method
I). The heating of the molecular ions heats the Be$^{+}$ ions sympathetically
and spectrally broadens their 313 nm absorption line. This leads to
a substantial increase of the cooling-laser stimulated atomic fluorescence,
due to the laser's relatively large detuning from atomic resonance.
The fluorescence signal level is indicative of the initial number
of HD$^{+}$ in the ion ensemble. (iii) For a duration $T_{c}=35\,$s
the rotational cooling takes place, after which the 2.7 \textmu{}m
rotational cooling laser is blocked (not the 5.5 \textmu{}m laser),
(iv) the REMPD lasers are unblocked and simultaneously the THz frequency
scan is initiated. The resulting molecule loss reduces the heating
and thus the atomic fluorescence signal. (v) The change in fluorescence
as a function of time is followed until the signal essentially reaches
the level in absence of molecules. (Fig.~\ref{fig: HD+ ion number reduction}
displays the first 60 s only). The secular excitation is kept on all
the time. This concludes acquisition of one data point. 

In an alternative measurement mode (method II), the secular scan activated
in step (ii) to obtain a normalization signal is turned off after
a few seconds while the rotational cooling continues. At the end of
step (iii) both rotational cooling lasers are blocked. In step (iv),
the THz source and the REMPD lasers are turned on only for 3 s. Immediately
afterward, in step (v) the secular excitation is turned on again and
the reduced fluorescence level is recorded during a few seconds. The
ratio of the two fluorescence levels defines our signal, and gives
approximately the relative decrease in HD\textsuperscript{+} number
after REMPD.

At the end of either procedure, residual HD\textsuperscript{+} and
product ions are removed from the trap by applying the following cleaning
procedure. (In method II, we do not {}``re-use'' the remaining molecules
as we prefer to excite molecular samples prepared in the same way
each time.) The cooling laser is detuned by a few 100 MHz to the red
of the atomic cooling transition, causing melting of the crystal into
a liquid. A secular excitation frequency scan covering the frequencies
of HD\textsuperscript{+} and lighter ions is turned on. The cooling
laser is briefly blocked and unblocked several times. Light ions are
thereby ejected from the trap. The secular excitation is turned off
and the system is ready for a new molecule loading.

\section{Results and Discussion}

\begin{figure}
\centering{}\includegraphics[bb=0bp 0bp 524bp 354bp,clip,scale=0.7]{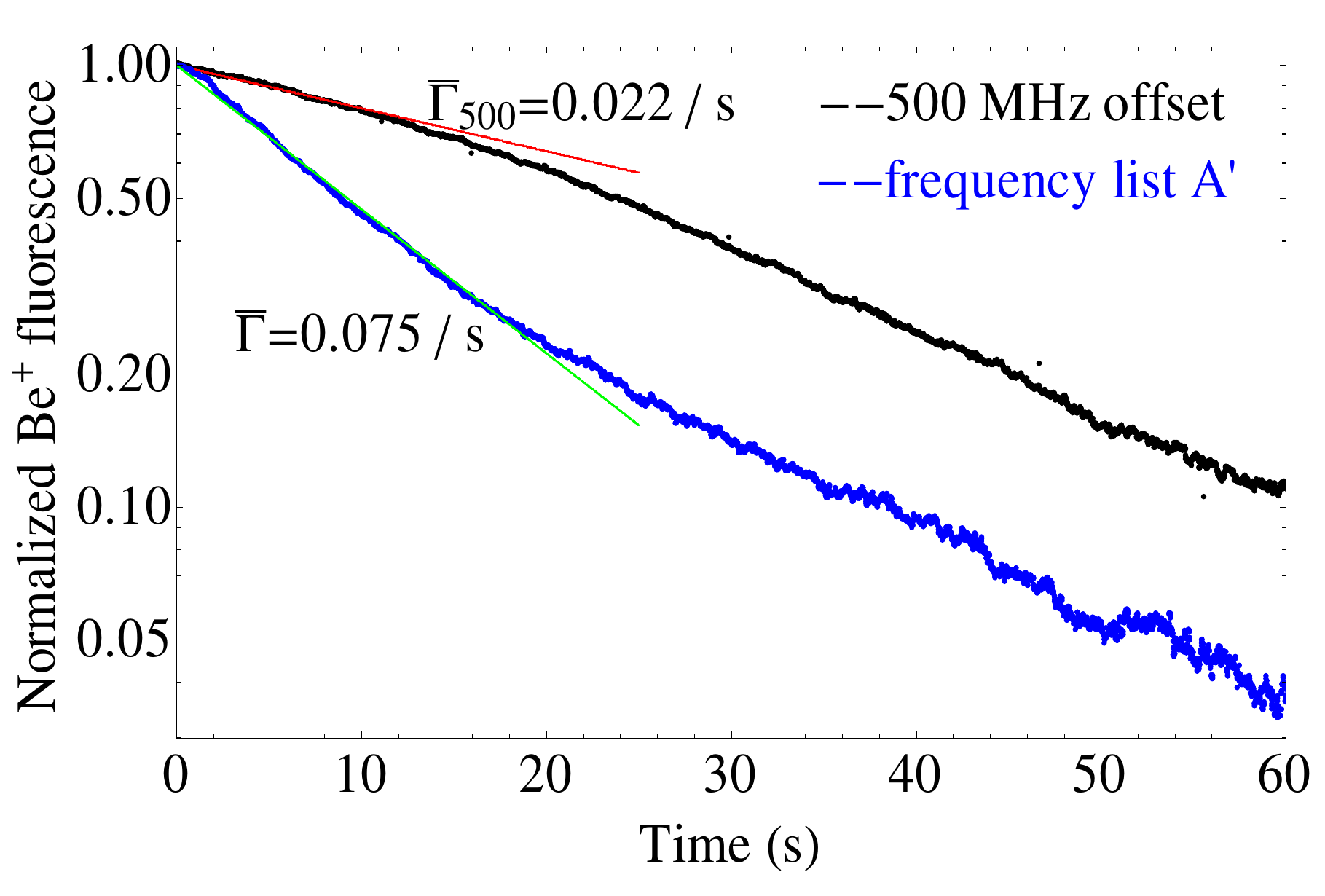}\caption{\label{fig: HD+ ion number reduction}\textbf{Atomic fluorescence
signal during continuous secular excitation of the HD\textsuperscript{+}ions
(method I, liquid state). }The time axis starts after the 2.7 \textmu{}m
rotational cooling laser is turned off and the two REMPD lasers and
THz radiation are turned on. The upper (black) trace, where the THz
radiation is detuned from resonance, shows the molecular ion number
decay due mainly to the effect of black-body-radiation-induced rotational
excitation. Lower (blue) trace: THz radiation is on-resonance. Each
trace is the average of 10 individual decays. The lines are exponential
fits to the first 10 s of the data. }
\end{figure}

The first set of measurements was taken in the liquid state and using
method I. Figure~\ref{fig: HD+ ion number reduction} shows two atomic
fluorescence traces. The upper (black, {}``background'') trace was
obtained with the THz wave frequency detuned by 500 MHz from $f_{0,th}$,
a value where no transition line exists. The REMPD only dissociates
molecules in the $(v=0,\, N=1)$ level. This level has initially been
depopulated by the 2.7 \textmu{}m rotational cooling laser, which
is favorable for the purpose of the spectroscopy. But as soon as the
spectroscopy phase starts, the level receives population not only
by THz rotational excitation (if the frequency is near resonance),
but also by BBR induced excitation from all hyperfine states of the
ground ro-vibrational level (rate ca. 0.09/s). In addition, population
still present in the $(v=0,\, N=2)$ level or reaching it from higher-lying
rotational levels is transferred into $(v=0,\, N=1)$ by BBR-stimulated
emission (rate ca. 0.12/s) and spontaneous emission (rate ca. 0.06/s).
Therefore, a REMPD-induced molecule loss is always present.

A rate equation simulation yields a molecule number decay rate that
depends on REMPD laser intensities and reaches ca. 0.075/s in the
limit of very high intensities. After partial optimization of the
UV laser alignment onto the ion crystal and therefore maximization
of its intensity, we observed values of ca. 0.04/s at 25 s after REMPD
laser turn-on. The value of the rate at this time rather than at 0
s is considered since at 0 s the number of molecules present is larger
and this may lead to some saturation of the secular excitation signal.
We explain the difference compared to the theoretical maximum by the
actually available laser power and possibly imperfect REMPD laser
beam overlap. 

The lower (blue) trace represents the decay in presence of resonant
THz radiation, using frequency list A'. We observe a large difference
in the initial rate of signal decrease as compared to the background
trace. Note that the decay occurring in presence of resonant THz radiation
also contains the background decay. 

\begin{figure}
\begin{centering}
\includegraphics[scale=0.35]{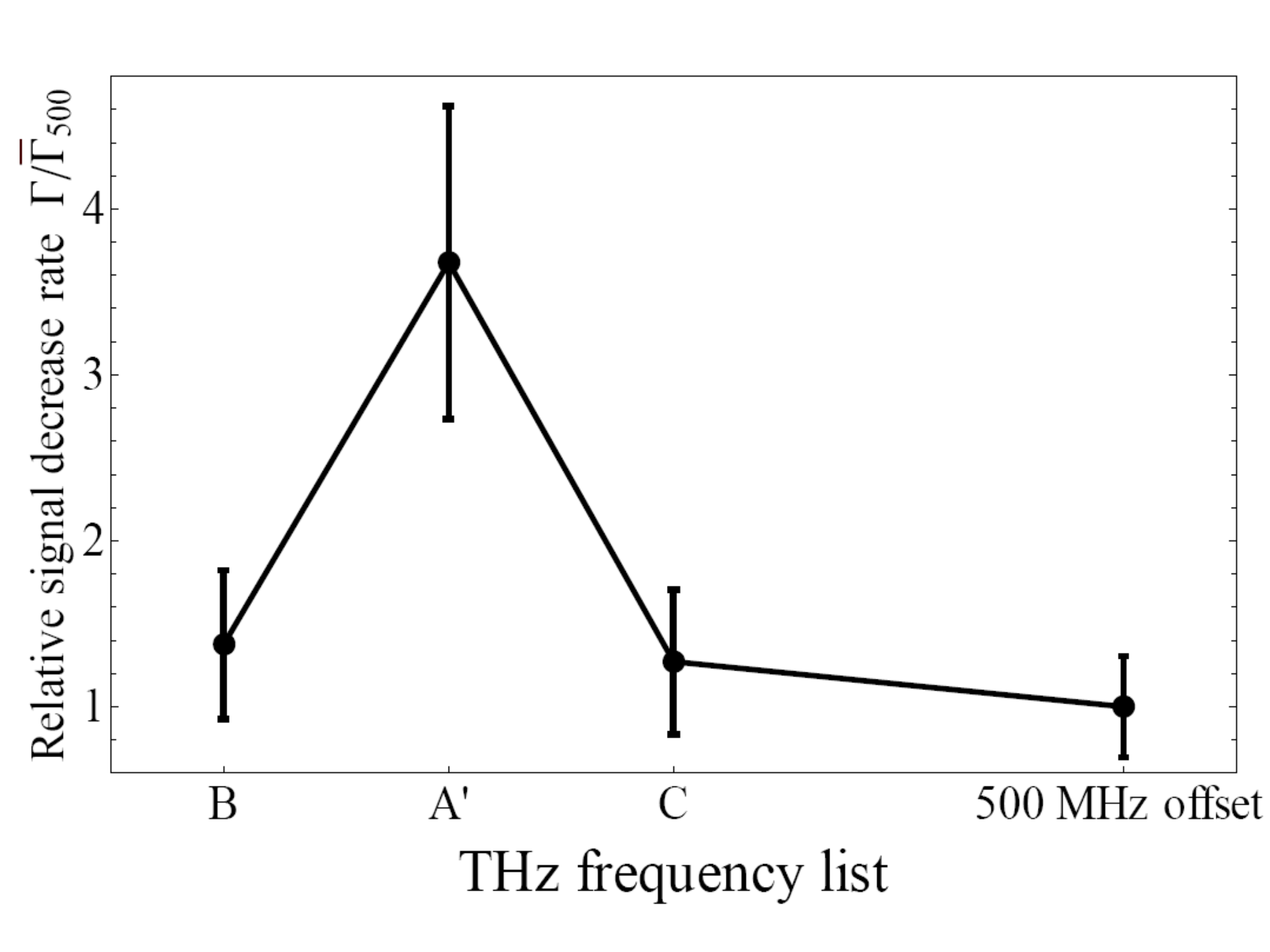}
\par\end{centering}

\begin{centering}
\includegraphics[scale=0.5]{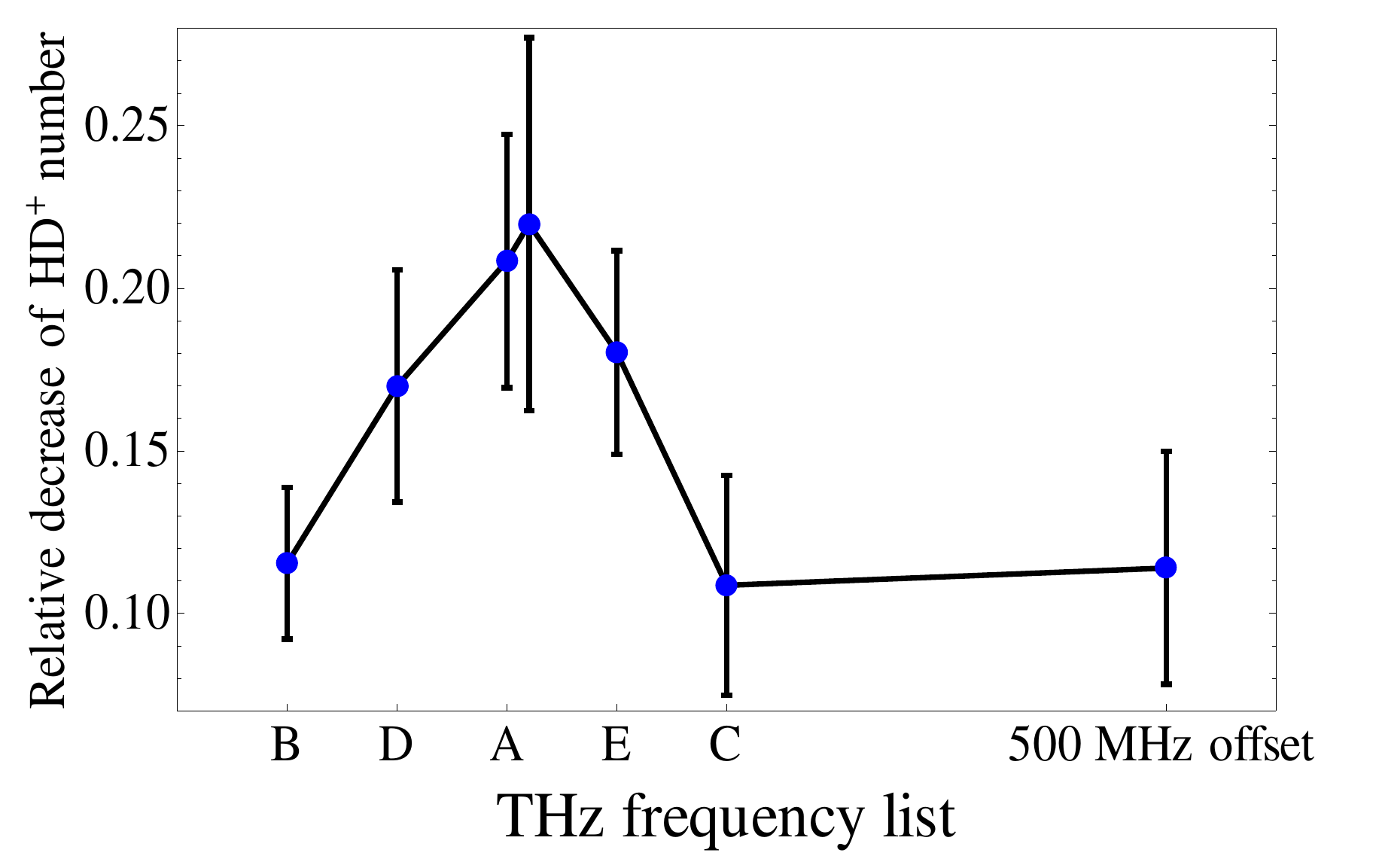}
\par\end{centering}

\caption{\label{fig: Spectrum}\textbf{Frequency dependence of the rotational
excitation.} Top: in the liquid state, at ca. 100 - 200 mK, using
method I. $\bar{\Gamma}_{500}$ is the average decay rate when the
THz radiation frequency is detuned by 500 MHz. Each data point results
from 9 individual decays. \protect \\
Bottom: in the crystallized state, at 10 - 15 mK, using method II.
Each data point represents the mean of 9 or 10 measurements. The two
close points were taken with the same list A on different days and
are shown separated for clarity. The error bars in both plots show
the standard deviations of the data, not of the mean. The lines are
guides for the eye.}
\end{figure}

We also performed measurements with the two frequency lists B, C,
for which the frequencies were detuned from those of list A' by different
amounts for the four hyperfine states. As Fig.\,\ref{fig: Spectrum}
shows, the decay rates do not differ significantly from the background
decay rates, and we can therefore deduce an upper limit for the magnetic
field of 1.5 G, as the influence of the Doppler width is not significant
here. 

A second set of measurements was taken with method II, see Fig.~\ref{fig: Spectrum}.
Here, the THz radiation is applied only when the ion ensemble is well-crystallized,
at temperatures of ca. 10 - 15 mK. Data points were taken alternately
at 500 MHz detuning, with list A, with list B (D), and with list C
(E). Again, a finite (background) signal is observed when the THz
radiation is far detuned, since BBR excites the rotational transition
significantly on the used timescale of 3~s. Irradiation with the
frequency list A provides a clear signal that rotational excitation
induced by THz radiation takes place. We find again no significant
difference in the rotational excitation efficiency for the detunings
of lists B and C, compared to the background measurement. However,
reducing the detunings to half the values (lists D, E), shows an increase
of signal. This increase can be explained by the presence of a magnetic
field of ca. 1 G or a Doppler width of several 100 kHz, or a combination
of both. However, the Doppler width is at most 70 kHz under the operating
conditions, therefore we conclude that we observed the effect of the
magnetic field on the hyperfine transition frequencies

\section{Conclusion}

We observed, for the first time, a pure rotational excitation of a
sympathetically cooled molecular ion ensemble. In order to facilitate
the observation, we applied a scheme adapted to the particularities
of the apparatus. As the available detection scheme is a destructive
one that employs photodissociation of the rotationally excited molecules,
a new molecular ion loading cycle has to be implemented for each data
point and the data acquisition rate is very low. The number of ions
sympathetically cooled is also small. Therefore, the preparation of
a significant fraction (ca. 70\%) of the molecular ions in the lower
spectroscopic level was essential. Even so, the detection of a rotational
transition originating from a single hyperfine state has too low a
signal-to-noise ratio. This is due to the concurrent process of black-body-radiation-induced
rotational excitation, which yields a finite background signal (decay
rate) in connection with the photodissociation. 

We therefore applied THz radiation at four frequencies that nominally
excite the four hyperfine states in the lower spectroscopic level.
This allowed a clear observation of the rotational excitation, using
two different methods. By applying THz radiation detuned from the
nominal resonance frequencies by amounts varying from 0.16 to 0.9
MHz for the four frequencies, we found a significantly reduced, but
still observable excitation. This can be explained by the presence
of a magnetic field with values up to approximately 1 G. In comparison,
the contribution of ion motion to the linewidth when the ions are
at ca. 10 mK and well-confined is negligible. 

There is a strong motivation for further development of the method
demonstrated here, since a resolution of the hyperfine structure of
HD\textsuperscript{+}and an accurate measurement of the hyperfine
transition frequencies would represent a significant test of the \textit{ab-initio}
calculations of this molecule. Possible improvements are the application
of hyperfine state preparation techniques, recently demonstrated \cite{Bressel 2011},
and accurate control of the magnetic field in the trap region.

Finally, it is useful to consider the extension of this work to other
molecular species. These can be characterized by their mass and their
rotational constant, which are to a certain extent related. For many
species, the masses will be significantly larger and the rotational
constants significantly smaller than in the case of HD\textsuperscript{+}.
The smaller rotational constant will lead to a significantly smaller
transition frequency, in the microwave regime. It is likely that the
Lamb-Dicke regime will then be effective, in particular if the microwave
propagation direction is along the narrow width of the molecular ion
ensemble, and Doppler broadening would be absent altogether. The smaller
rotational constant will also lead to a smaller black-body-radiation
excitation rate (at 300 K), which is very favorable, since it will
make possible REMPD with near-zero background. The experimentally
demonstrated fraction of molecules in the ground state obtained by
applying rotational cooling on such heavier molecules is so far significantly
below the level used here on HD\textsuperscript{+}\cite{Staanum rot. cooling},
but simulations \cite{Supplemental material in Schneider et al.}
show that similar levels should be achievable with appropriate laser
cooling schemes and laser systems. Hyperfine structure and Zeeman
shifts coefficients will be molecule-specific. Molecular ions in an
electronic spin singlet state are particularly interesting, as they
would have a reduced number of hyperfine states, Zeeman sub-states
and much reduced linear Zeeman shift coefficients, simplifying and
narrowing the spectrum. Thus, the extension of rotational spectroscopy
of sympathetically cooled molecular ions to other species appears
very promising.

\textit{Note added: With an improved alignment of the UV laser, an
increase of the decay rate in absence of THz radiation to ca. 0.060/s
was observed.}

\section*{Acknowledgment }

This work was funded by the DFG (project Schi 431/11-1) and by an
equipment grant of the Heinrich-Heine-Universität. We thank F. Lewen
for loan of equipment and V. Korobov for communication of unpublished
results. We are indebted to B. Roth and T. Schneider for their contributions
in the initial phase of this study.

\end{document}